\documentclass[aps,prb,superscriptaddress,twocolumn,showpacs,showkeys,floatfix]{revtex4-1}
\pdfoutput=1
\usepackage{amsmath}
\usepackage{bm}
\usepackage{amssymb}
\usepackage{graphicx}
\usepackage{epsfig}
\usepackage{color}
\usepackage{braket}
\usepackage{lipsum}
\begin{document}

\title{Towards nanoscale multiplexing with parity-time symmetric plasmonic coaxial waveguides}
\author{Hadiseh Alaeian}
\thanks{H. Alaeian and B. Baum contributed equally}
\affiliation{Institut f\"ur Angewandte Physik, University of Bonn, 53115 Bonn, Germany}
\affiliation{Department of Electrical Engineering, Stanford University, Stanford, California 94305, USA}
\affiliation{Department of Materials Science and Engineering, Stanford University, Stanford, California 94305, USA}
\author{Brian Baum}
\thanks{H. Alaeian and B. Baum contributed equally}
\affiliation{Department of Materials Science and Engineering, Stanford University, Stanford, California 94305, USA}
\author{Vladan Jankovic}
\affiliation{Northrop Grumman Aerospace Systems, One Space Park, Redondo Beach, California 90278, USA}
\author{Mark Lawrence}
\affiliation{Department of Materials Science and Engineering, Stanford University, Stanford, California 94305, USA}
\author{Jennifer A. Dionne}
\affiliation{Department of Materials Science and Engineering, Stanford University, Stanford, California 94305, USA}
\date{\today}

\begin{abstract}

We theoretically investigate a nanoscale mode-division multiplexing scheme based on parity-time ($\mathcal{PT}$) symmetric coaxial plasmonic waveguides. Coaxial waveguides support paired degenerate modes corresponding to distinct orbital angular momentum states. $\mathcal{PT}$ symmetric inclusions of gain and loss break the degeneracy of the paired modes and create new hybrid modes without orbital angular momentum. This process can be made thresholdless by matching the mode order with the number of gain and loss sections within the coaxial ring. Using both a Hamiltonian formulation and degenerate perturbation theory, we show how the wavevectors and fields evolve with increased loss/gain and derive sufficient conditions for thresholdless transitions. As a multiplexing filter, this $\mathcal{PT}$ symmetric coaxial waveguide could help double density rates in on-chip nanophotonic networks. 

\end{abstract}

\maketitle 
\section{introduction}

The increasing size and sophistication of supercomputers and server farms have necessitated the development of new high data-rate, high efficiency, and compact devices for the transfer of information across a variety of distances from cross-chip to cross-warehouse. Compared to electronic interconnects, a single optical interconnect can carry extremely high data rates via a number of alternative and complementary multiplexing schemes. For example, IBM's silicon photonics chip enables data-rates up to 100 Gb/s using wavelength-division multiplexing with a demonstrated error-free operation at 32 GB/s.~\cite{Gill:2015tp} The inclusion of multiple systems of multiplexing, e.g., time and mode-division multiplexing, could offer an additional significant increase to the data flux per optical channel, since multiplexing schemes are multiplicative in data-rate. 

While wavelength- and time-division multiplexing are the most mature technologies for increasing data-rates, mode-division multiplexing (MDM) has gained increasing attention since its inception in the 1980s.~\cite{Berdague:1982ws} With this scheme, information is encoded within specific spatial eigenmodes. Terabit-scale data rates have already been demonstrated with MDM using two modes with different orbital angular momentum (OAM) modes for ten wavelengths.~\cite{Bozinovic:2013ch} However, these demonstrations of MDM have mostly relied on free space or fiber transmission and bulky multiplexing/demultiplexing components.~\cite{Bai:12,Boffi:2013kr,Huang:2015ht} On-chip MDM is possible, but still requires relatively large optical components to filter the different modes.~\cite{Luo:2014gy,Dorin:2014fp,Stern:2015ha} 

Plasmonic waveguides offer substantial field confinement and concentration with modest mode propagation lengths.~\cite{Leosson:2006uf,Gramotnev:2010ji,DeLeon:2010cz,Berini:2009fs} Plasmonic coaxial metal-insultator-metal structures have shown promise as field concentrators~\cite{Baida:2006fi, Orbons:2007er,Min09,Vesseur:2011wk,Saleh:2012bz} for nanoscale optical waveguides and on-chip thresholdless lasers.~\cite{Khajavikhan2012,Nezhad10} As waveguides, these structures exhibit large modal gain,~\cite{Saleh:2012em} which can mitigate the Ohmic losses generally associated with this high level of confinement. Orthogonal plasmonic modes (i.e., non-mixing modes) that share the same effective index can multiply on-chip data rates by the number of modes, provided a means of separating and individually addressing the modes is devised. In this work, we propose a scheme for multiplexing with plasmonic coaxial waveguides based on parity-time ($\mathcal{PT}$) symmetry breaking. Plasmonic coaxial waveguides possess a number of paired degenerate modes, corresponding to clockwise (CW) and counter-clockwise (CCW) OAM states, which undergo useful transformations with the introduction of $\mathcal{PT}$ symmetry. 

By designing a $\mathcal{PT}$ symmetric azimuthally periodic perturbation within the waveguide, one can lift the OAM mode degeneracy, and create new hybrid modes. These modes lose their OAM and become amplifying or attenuating. This effect could therefore be used as a multiplexing filter, which could feed a selected data stream into a specific mode of a final plasmonic coaxial channel. Control over individual modes is achieved by perturbing the insulator within the coaxial waveguide with rotationally periodic gain and loss media with the same real refractive index and equal magnitude imaginary index (which is the necessary condition to possess $\mathcal{PT}$ symmetry in optical systems). 

In one-dimensional $\mathcal{PT}$ symmetric systems, as the amount of gain and loss increases in the system, an exceptional point (EP) is reached beyond which the eigenvalues and eigenvectors of the system markedly shift, and modes will pair off and become complex conjugates (same real refractive index, but equal and opposite imaginary index) of each other. This phenomenon has been used to great effect to design waveguides,~\cite{Ge:2014fd, Vysloukh:2014kw, Huang:2014ui, Yu:2012fe,Barashenkov:2013jq,Scott:2012kl,Lupu:2013cj,Guo:2009hd,Ruter:2010dy}  resonators,~\cite{Bender:2013eg, Hodaei:2014kr, Longhi:2014gp, Peng:2014kl} and periodic structures exhibiting asymmetric transmission.~\cite{Kulishov:2013fe,Regensburger:2012jm}  Further, low threshold single mode lasing has been demonstrated for micron-scale $\mathcal{PT}$ symmetric ring-resonators.~\cite{Feng:2014gg} In our system, the well-defined and azimuthally-localized field distributions in coaxial geometries make them an ideal candidate for the exploration of more sophisticated $\mathcal{PT}$ symmetric devices as a means of mode filtering and shaping. 

In addition to the more common $\mathcal{PT}$ symmetric EP behavior, we show it is possible to break the modal degeneracy with arbitrarily small $\mathcal{PT}$ symmetric perturbations in our plasmonic coaxial waveguide, thereby enabling the differentiation between a pair of modes that share the same real part of the wavevector. We also show how the gain/loss arrangements can be used to design new hybrid modes from the modes of a uniform coaxial waveguide. Using both a Hamiltonian formulation and degenerate perturbation theory, we investigate the evolution of the modes as a function of the gain/loss value and the number of gain/loss sections in the waveguide; we obtain the sufficient conditions to achieve both thresholdless $\mathcal{PT}$ symmetry breaking as well as classic EP behavior. In addition to mode selection for MDM, the results of this study in general shed light on the mode morphology and $\mathcal{PT}$ symmetry breaking in systems with degeneracy. 

\section{theoretical formulation}

\begin{figure}[b]
\includegraphics[scale=1]{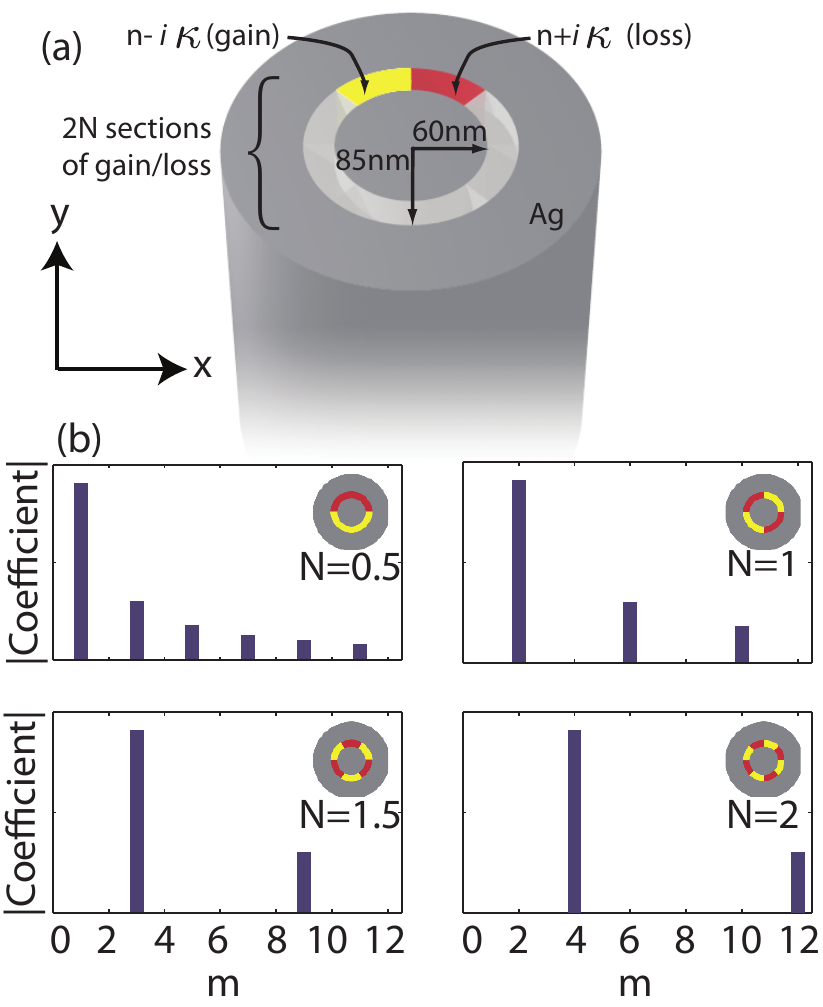}
\caption{\label{Fig1} (a) Schematic of the $\mathcal{PT}$ symmetric coaxial waveguide where the channel is perturbed with alternating sections of the gain(yellow) and loss(red). (b) The Fourier coefficients defining $\kappa$ for each distribution $N=0.5, 1, 1.5, 2$. Due to the symmetry present, only positive-order coefficients are shown.}
\end{figure}

As shown in Fig.~\ref{Fig1}(a), our structure is a three layer coaxial waveguide consisting of a dielectric ring and silver core embedded in a silver cladding. The silver core has a radius of 60\,nm and the permittivity is described with a Drude model as $\epsilon_{Ag}=1-(\frac{\omega_p}{\omega})^2$, where $\omega_p=8.85\times 10^{15}$ (Discussion of a more realistic metal including Ohmic loss is given in Appendix A). The dielectric channel is 25\,nm thick and characterized by a real refractive index of $n=1.5$. The channel is filled with alternating sections of gain and loss, which can be represented as:

\begin{equation}\label{perturbing index}
\Delta n=
\begin{cases}
n_{gain}=-i\kappa  & l\pi/N\leq\phi\leq(l+1/2)\pi/N\\
n_{loss}=+i\kappa  & (l+1/2)\pi/N\leq\phi\leq(l+1)\pi/N\\
\end{cases}.
\end{equation}
Here, $\kappa$ is the magnitude of the gain/loss, $2N$ is the number of loss (or gain) segments, and $l$ is an integer spanning  $0,1,\cdots 2N-1$. The presence of gain and loss makes the system non-Hermitian, so $\kappa$ can be considered the non-Hermiticity parameter. The refractive index profile of the coaxial waveguide satisfies the $\mathcal{PT}$ symmetry condition [$n(y)=n(-y)$, $\kappa(y)=-\kappa(-y)$]. While a closed-form solution to this $\mathcal{PT}$ coaxial waveguide does not exist, a Hamiltonian formulation~\cite{Alaeian:2014eb}  and degenerate perturbation theory can be used to investigate the modal properties.

Due to the axial symmetry of the waveguide, the modes vary azimuthally as $e^{im\phi}$; in other words, they have well-defined angular momenta parametrized by $m$. For a periodic refractive index distribution as in Eq.~\ref{perturbing index}, the index can also be expanded in the harmonic basis of $e^{im\phi}$. The overall symmetry of the waveguide's cross section can be categorized based on whether $N$ is an integer or half integer:

$N$ (\emph{half-integer}): the distribution possesses anti-symmetry [$\kappa(x,y)=-\kappa(-x,-y)$], so \emph{even}-order Fourier coefficients vanish.

$N$ (\emph{integer}): the distribution has inversion symmetry [$\kappa(x,y)=\kappa(-x,-y)$], so \emph{odd}-order Fourier coefficients vanish.

Figure~\ref{Fig1}(b) shows the Fourier coefficients for the distribution of $\kappa$ of four different $\mathcal{PT}$ symmetric distributions. In each case the inset shows a cross section of each waveguide. Due to the symmetry condition of $C_m = C_{-m}^*$, only positive-order Fourier coefficients (i.e., $m\geq 0$) are shown in each case. The transverse electric and magnetic fields, $\vec{E}_{t}$ and $\vec{H}_{t}$, and corresponding propagation constants $\beta$ of the waveguide modes are determined via:  

\begin{equation}\label{Hamiltonian form}
\hat{H}\begin{bmatrix}
\vec{E}_t\\\vec{H}_t
\end{bmatrix}=\beta\begin{bmatrix}
0 & -\hat{z}\times\\ \hat{z}\times & 0
\end{bmatrix}
\begin{bmatrix}
\vec{E}_t\\\vec{H}_t
\end{bmatrix}.
\end{equation}
Here, the superscript \emph{t} refers to the transverse components of the fields and $\hat{z}$ is the direction of propagation. The Hamiltonian $\hat{H}$ depends on the waveguide geometry, as well as the material properties, and is given by:

\begin{widetext}
\begin{equation}
\hat{H}=\begin{bmatrix}
\omega \epsilon_0 \epsilon-\frac{1}{\omega\mu_0}\bigtriangledown_t\times(\hat{z}(\hat{z}\cdot\bigtriangledown_t\times)) & 0\\
0 & \omega\mu_0-\frac{1}{\omega\epsilon_0}\bigtriangledown_t\times(\hat{z}\frac{1}{\epsilon}(\hat{z}\cdot\bigtriangledown_t\times))
\end{bmatrix},
\end{equation}
\end{widetext}
where $\epsilon_0$ and $\mu_0$ are the permittivities and permeabilities of free space, $\epsilon$ is the material permittivity, and $\omega$ is the optical angular frequency. By choosing a complete basis $\ket{\vec{F}_n}$, the matrix elements of the Hamiltonian $\hat{H}$ can be determined as:

\begin{equation}\label{equivalent matrix}
\begin{aligned}
&H_{mn}=\braket{\vec{F}_m|\hat{H}|\vec{F}_n}\\
&=\omega\epsilon_0\int\limits_{WG}\epsilon(r)\vec{E}^*_{mt}\cdot\vec{E}_{nt} ds + \omega\mu_0\int\limits_{WG}\vec{H}^*_{mt}\cdot\vec{H}_{nt} ds \\
&-\omega\epsilon_0\int\limits_{WG}\epsilon(r)^*{E}^*_{mz} E_{nz} ds-\omega\mu_0\int\limits_{WG}H^*_{mz} H_{nz} ds,
\end{aligned}
\end{equation}
where subscripts $m$ and $n$ refer to the $m^{th}$ and $n^{th}$ modes respectively, subscript $z$ denotes the longitudinal direction, and the integral $\int\limits_{WG} ds$ indicates an integration over the cross section of the waveguide. From this last equation it can be inferred that $\hat{H}$ is Hermitian if and only if $\epsilon(r)$ is real, i.e., all the materials are lossless. 

\begin{figure}
\includegraphics[scale=1]{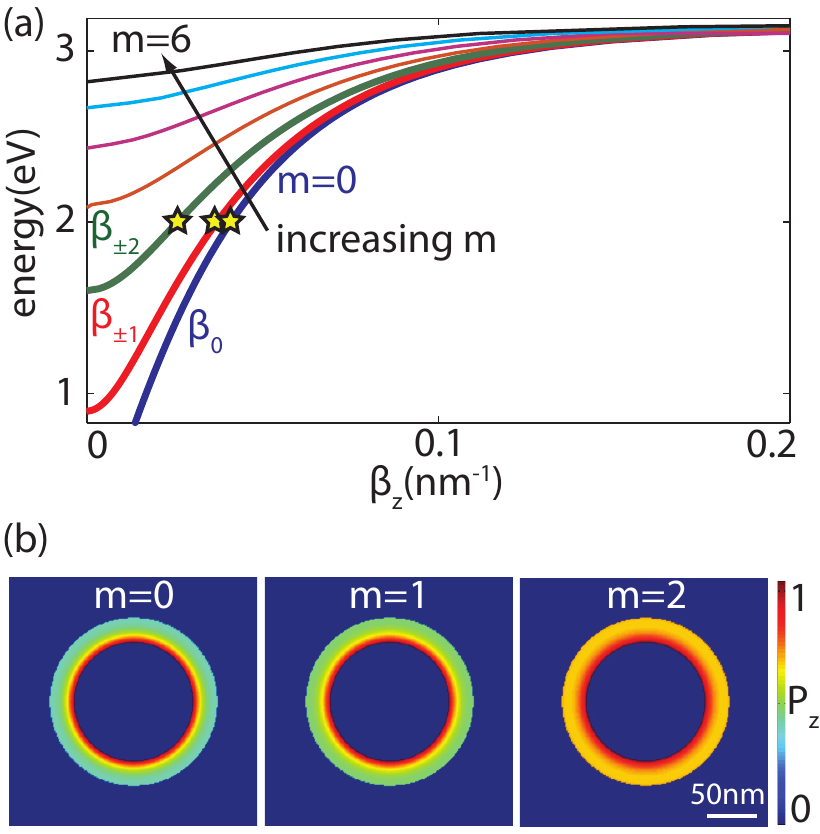}
\caption{\label{Fig2} (a) Modes of the uniform coaxial waveguide numbered according to the azimuthal order \emph{m}. For the rest of the paper, we focus on $E= 2\,$eV, which is marked on the first 3 modes as stars. (b) Distribution of the Poynting vector component $P_z$ for $m=0,1,2$ at $E= 2\,$eV.}
\end{figure}

\begin{figure}[b]
\includegraphics[scale=1]{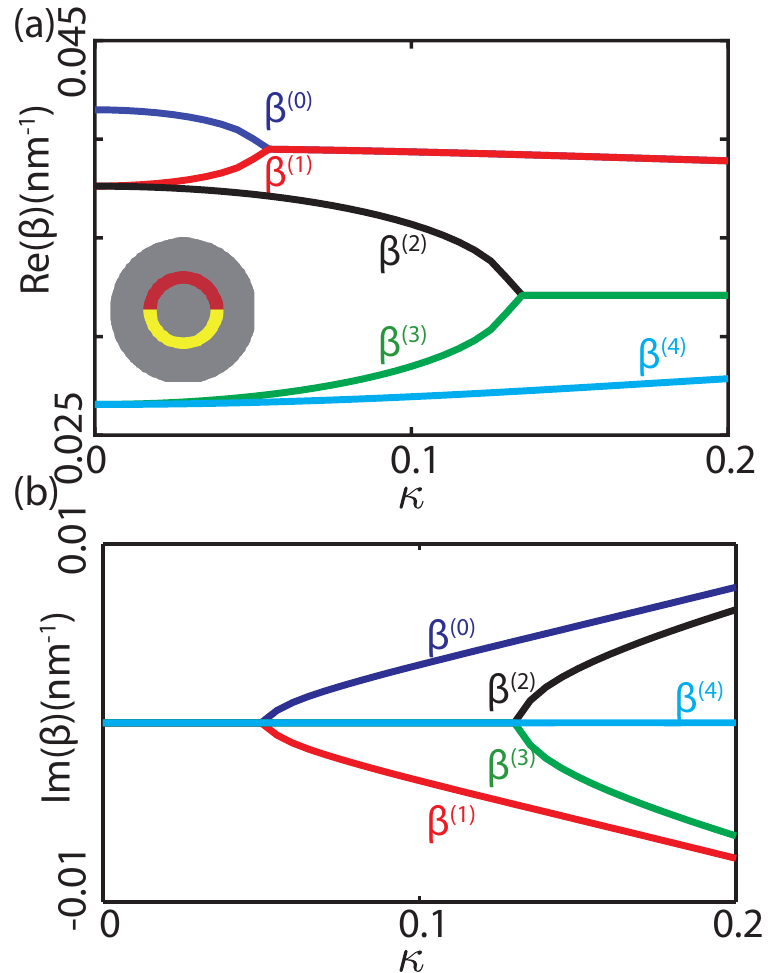}
\caption{\label{Fig3} (a)Real and (b) imaginary parts of the propagation constant of the five lowest order modes of a $\mathcal{PT}$ symmetric coaxial waveguide when $N=0.5$ at $E=2\,$eV. The inset shows the schematic of the waveguide cross section.}
\end{figure}

For $\kappa \neq 0$, the Hamiltonian can be written as $\hat{H}=\hat{H}_R+i\hat{H}_I$. When $\kappa\ll n$, $\hat{H}_R$ of the $\mathcal{PT}$ symmetric and the uniform waveguide are nearly equal---hence $\hat{H}_R$ is diagonal in the basis of the homogenous waveguide modes. Using Eq.~\ref{equivalent matrix} we have:

\begin{equation}\label{H-matrix}
\begin{split}
H_{mn} =\braket{\vec{F}_m|\hat{H}_R+i\hat{H}_I|\vec{F}_n} &=\braket{\vec{F}_m|\hat{H}_R|\vec{F}_n}+i\braket{\vec{F}_m|\hat{H}_I|\vec{F}_n}\\
&= \beta_n \delta_{mn}+i\braket{\vec{F}_m|\hat{H}_I|\vec{F}_n}.
\end{split}
\end{equation}

The matrix elements of the perturbing Hamiltonian are determined as:

\begin{equation}\label{HI-matrix}
H_{I_{mn}}=\braket{\vec{F}_m|\hat{H}_I|\vec{F}_n}=\omega\epsilon_0\int\limits_{WG}\epsilon_i(r)\vec{E}^*_m\cdot \vec{E}_n ds.
\end{equation}
Here $\epsilon_i(r)$ is the imaginary part of the dielectric constant spatially modulated as in Eq.~\ref{perturbing index}. ~\footnote{In Eq.~\ref{equivalent matrix}, $\vec{E}_m$ includes both of the transverse($\vec{E}_{mt}$) and longitudinal ($E_{mz}$) components.} Considering the azimuthal variation of the modes in a uniform waveguide, the above equation can be simplified to: 

\begin{widetext}
\begin{equation}\label{HI-matrix2}
\begin{split}
H_{I_{mn}}=\braket{\vec{F}_m|\hat{H}_I|\vec{F}_n} =\omega\epsilon_0\int\limits_{R_{in}}^{R_{out}}dr r  \vec{R}_m^*(r)\cdot\vec{R}_n(r)\int\limits_{0}^{2\pi}d\phi \epsilon_i(\phi)^{i(n-m)\phi}
=2\pi\omega\epsilon_0C_{m-n}\int\limits_{R_{in}}^{R_{out}}dr r  \vec{R}_m^*(r)\cdot\vec{R}_n(r),
\end{split}
\end{equation}
\end{widetext}
where $\vec{R}(r)$ is the radial distribution of the modes and $C_{m-n}$ is the $(m-n)^{th}$ Fourier coefficient of the gain/loss arrangements in the channel given in Eq.~\ref{perturbing index}. Since $C_{m-n}$ describes the gain and loss, it is linearly proportional to $\kappa$. As shown in Fig.~\ref{Fig1}(b), the Fourier coefficients can be largely controlled via $N$. The modal properties of each case can be determined by solving for the eigenvalues and eigenfunctions of the $H$-matrix given by Eq.~\ref{H-matrix}.

\section{Dispersion and Fields}

We first apply this formalism to a coaxial waveguide with a uniform channel ($\kappa=0$). Figure~\ref{Fig2}(a) plots the dispersion for this coaxial structure, where mode orders up to $m=6$ are found below $3\,$eV. Note that the dispersions of all modes diverge for energies close to the surface resonance frequency of the silver-dielectric interface ($\omega_{sp}\approx 3.1$\,eV). The modes of the uniform waveguide ($\kappa=0$) form a complete set, so they can be used as a basis $\ket{\vec{F}_n}$ to find the modes of the $\mathcal{PT}$ symmetric waveguides of Fig.~\ref{Fig1}(b). Figure~\ref{Fig2}(b) shows the distribution of the longitudinal power $P_z$ of the $0^{th}$, $1^{st}$, and $2^{nd}$-order modes. Note that these modes all have azimuthally symmetric power distributions as a result of possessing well-defined angular momenta.

\begin{figure*}
\includegraphics[scale=1]{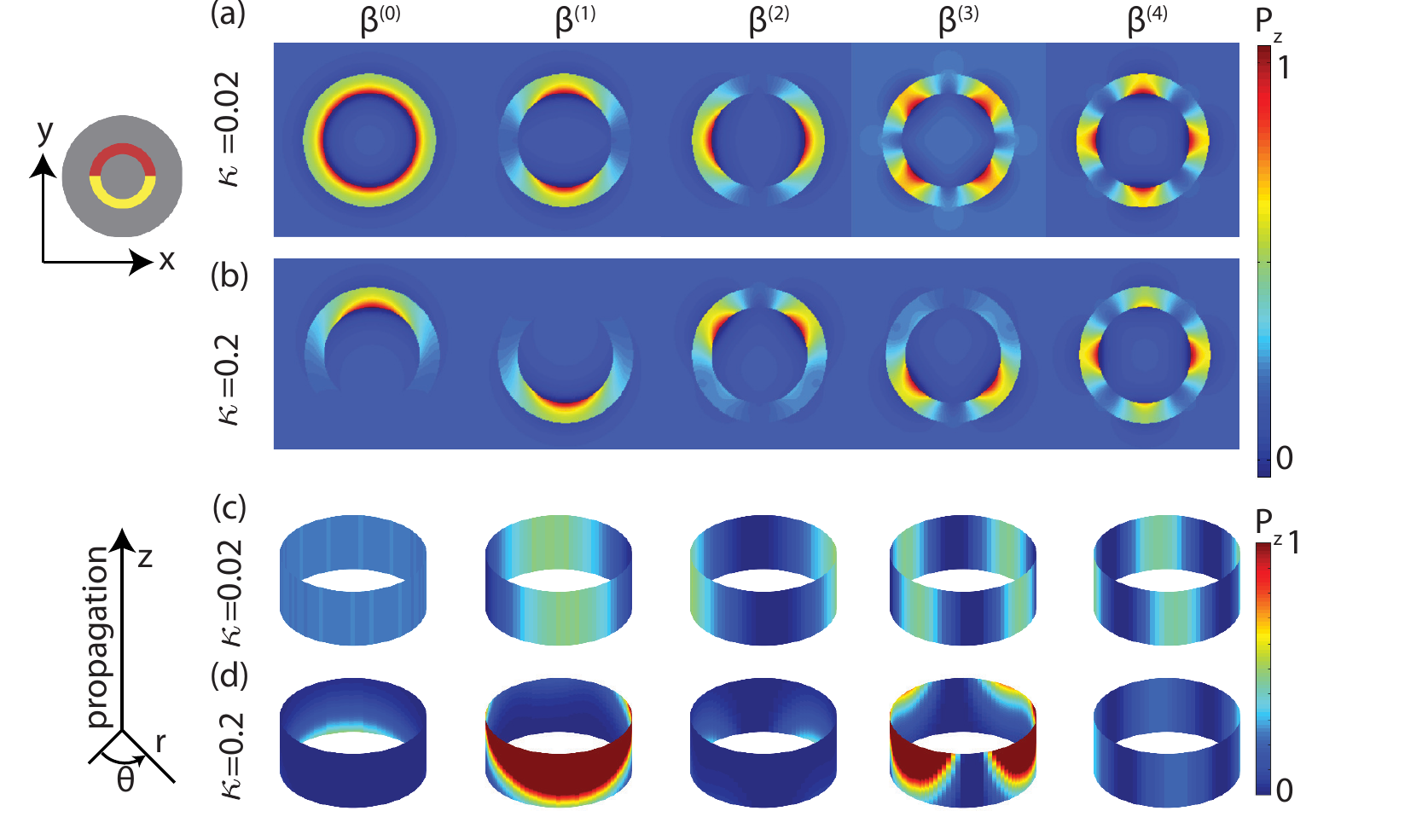}
\caption{\label{fig4}  (a) The out-of-plane component of the Poynting vector $P_z$ for the five lowest order modes of a $\mathcal{PT}$ symmetric coaxial waveguide, when $N=0.5$ at $E=2$\,eV is plotted for each mode of the perturbed waveguide at $\kappa=0.02$ and (b) $\kappa=0.2$. (c) and (d) show the magnitude of the Poynting vector along the propagation direction for the different modes for a slice in the ring 5\,nm from the core interface. The rings are rotated by $5^{\circ}$ for clarity.}
\end{figure*}

To investigate the effect of loss and gain inclusions on the modal properties, we consider a fixed energy, $E=2$\,eV. As seen in Fig.~\ref{Fig2}, the uniform waveguide supports five modes at this energy, namely $\beta_0$, $\beta_{\pm1}$, and $\beta_{\pm2}$. $\beta_{\pm}$ represents a pair of degenerate modes, corresponding to CW and CCW OAM. $E=2$\,eV therefore limits the total number of modes to a representative set. Utilizing these modes as the basis of expansion in Eq.~\ref{equivalent matrix}, one can investigate the mode morphology as a function of both $N$ and $\kappa$. Figure~\ref{Fig3} shows the modal properties of a coaxial waveguide with $N=0.5$ at $E=2$\,eV. Panels (a) and (b) show the variation of the real and imaginary part of the propagation constants of the five lowest order modes as a function of $\kappa$. To differentiate the new modes that appear when $\kappa\ne0$ from the $\kappa=0$ modes, superscript indexing has been used. At $\kappa=0$ in Fig.~\ref{Fig3}, all modes possess the same propagation constants as in Fig.~\ref{Fig2}, and the superscript notation eigenvalues can therefore be matched to the unperturbed subscript eigenvalues. The $\beta^{(0)}$-branch has the largest propagation constant and at $\kappa=0$ corresponds to the $m=0$ mode $\beta_{0}$. The degenerate pair $\beta_{\pm1}$ become $\beta^{(1)}$ and $\beta^{(2)}$ for $\kappa\ne0$. Similarly, the degenerate pair $\beta_{\pm2}$ become $\beta^{(3)}$ and $\beta^{(4)}$ for $\kappa\ne0$. 

As $\kappa$ increases, the real($\beta$) of degenerate modes separate from one another (i.e., $\beta^{(1)}$ from $\beta^{(2)}$, $\beta^{(3)}$ from $\beta^{(4)}$) and at $\kappa=0.05$, $\beta^{(0)}$ and $\beta^{(1)}$ reach an EP and real($\beta$) coalesce. $\beta^{(2)}$ and $\beta^{(3)}$ form a similar pair and coalesce at an EP at $\kappa=0.13$. As discussed in Appendix B, the propagation constants of these modes are always either real or complex conjugates of each other. To clarify this feature, we present the imaginary parts of the propagation constants in Fig.~\ref{Fig3}(b). Note that where the real wavevectors converge [Fig.~\ref{Fig3}(a)], the imaginary parts diverge. The expansion coefficients of these modes before and beyond their EPs are given in Appendix C.

Figure~\ref{fig4} shows the spatial distribution of the longitudinal component of the Poynting vector $P_z$ of the modes at small and large values of $\kappa$. For small values of non-Hermiticity and before all EPs ($\kappa=0.02$), the propagation constants of all the modes are real, since the power is symmetrically distributed in the gain and loss sections. As seen in Fig.~\ref{fig4}(a), the power distributions of all modes are symmetric with respect to the $x$-axis; however, in contrast to the uniform waveguide, the power is no longer azimuthally symmetric. When the non-Hermiticity factor is increased to a value beyond both EPs ($\kappa=0.2$), new complex conjugate modes are formed. These modes, on display in Fig.~\ref{fig4}(b), lose their symmetry with respect to the $x$-axis and are either localized mainly to the loss or gain half of the waveguide. The only mode of this set which does not lose its $x$-axis symmetry plane is $\beta^{(4)}$, which possesses no partner at $E=2$\,eV with which to form a complex conjugate pair. At higher values of $\kappa$ and for $E>2$\,eV, this mode would also eventually reach an EP and lose its symmetry. To better visualize the imaginary part of the propagation constants of these modes, Fig.~\ref{fig4}(c) and (d) show radial slices within the dielectric ring 5\,nm from the core interface. Before the EP, all modes show unchanging $P_z$ along the propagation direction. Beyond the EP, $P_z$ drastically increases for $\beta^{(1)}$ and $\beta^{(3)}$, and diminishes for $\beta^{(0)}$ and $\beta^{(2)}$.
 
\begin{figure}
\includegraphics[scale=1]{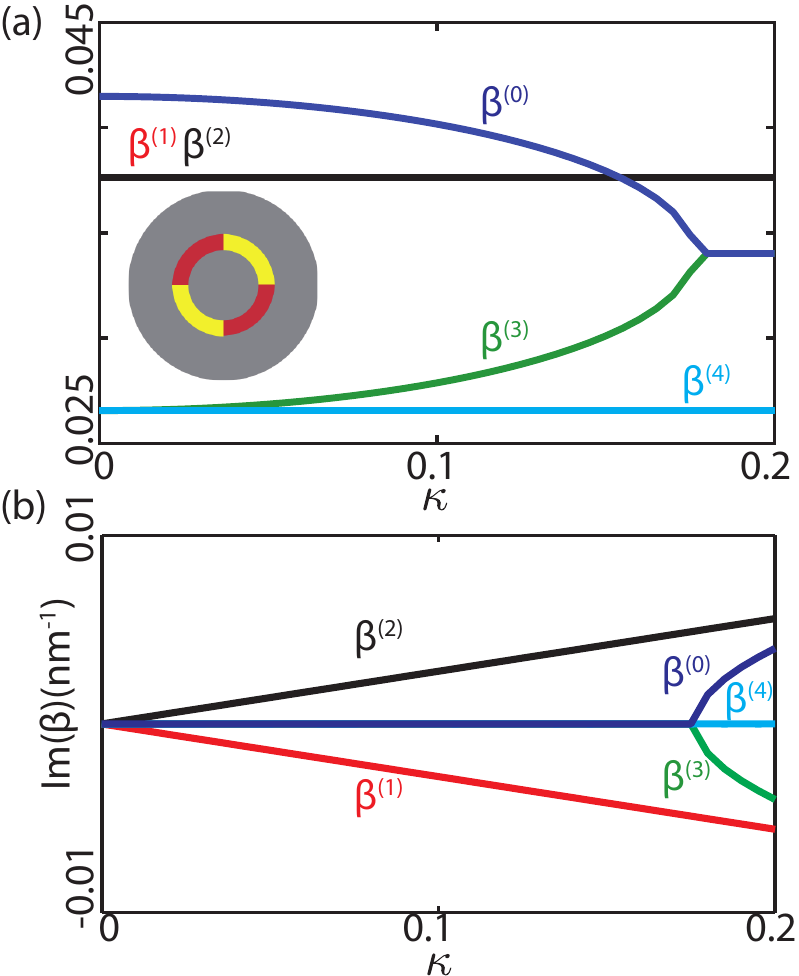}
\caption{\label{fig5}  (a)Real and (b) imaginary parts of the propagation constant of the five lowest order modes of a $\mathcal{PT}$ symmetric coaxial waveguide when $N=1$ at $E=2$\,eV. The inset shows the schematic of the waveguide cross section.}
\end{figure}

\begin{figure*}
\includegraphics[scale=1]{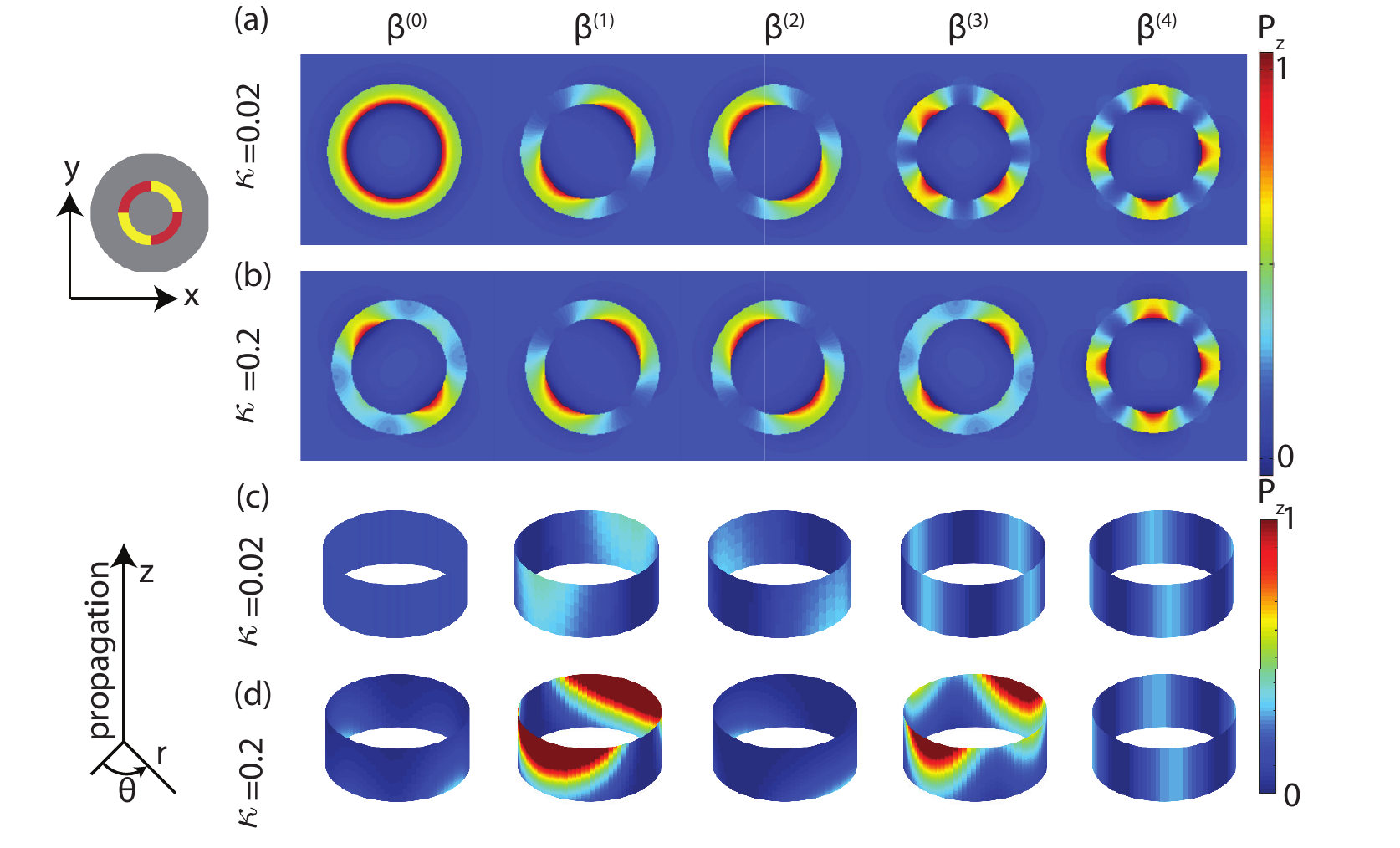}
\caption{\label{fig6}  (a) The out-of-plane component of the Poynting vector $P_z$ for the five lowest order modes of a $\mathcal{PT}$ symmetric coaxial waveguide, when $N=1$ at $E=2$\,eV is plotted for each mode of the perturbed waveguide at $\kappa=0.02$ and (b) $\kappa=0.2$. (c) and (d) show the magnitude of the Poynting vector along the propagation direction for the different modes for a slice in the ring 5\,nm from the core interface. The rings are rotated by $5^{\circ}$ for clarity.}
\end{figure*}

The results for $N=0.5$ are reminiscent of classical EP behavior in one-dimensional systems: namely, a finite value of $\kappa$ is required to induce mode coalescence and enter the broken-phase regime. However, because the coaxial waveguide supports degenerate modes, thresholdless behavior can be achieved provided the $\mathcal{PT}$ symmetry is engineered correctly.\cite{Ge:2014fd} Such behavior requires that the mode symmetry match the distribution of loss and gain---a condition that can be met when $N=1$, but which can also be satisfied for higher order modes with higher values of $N$, as will be discussed. The variations of the propagation constants as well as the corresponding power distributions for this structure are given in Fig.~\ref{fig5} and Fig.~\ref{fig6} for $E=2$\,eV. For this case, modes $\beta^{(1)}$ and $\beta^{(2)}$ possess a constant real propagation constant. The imaginary parts of these modes separate from each other for $\kappa\ne 0$ and therefore have no EP. We also witness a new pairing between $\beta^{(0)}$ and $\beta^{(3)}$, which reach an EP at $\kappa=0.175$. We note this EP occurs at a higher $\kappa$ than both the pairing of $\beta^{(0)}$--$\beta^{(1)}$ and $\beta^{(2)}$--$\beta^{(3)}$ for the $N=0.5$ coaxial waveguide because $\beta_0$--$\beta_{\pm2}$ have a greater separation than $\beta_0$--$\beta_{\pm1}$ and $\beta_{\pm1}$--$\beta_{\pm2}$. The mode with the smallest wavevector, $\beta^{(4)}$, remains unaffected by the inclusion of gain and loss and possesses a constant real propagation constant with no imaginary part.

Figure~\ref{fig6} shows the power distribution for this $N=1$ coaxial waveguide, again at $\kappa=0.02$ (a) and $0.2$ (b). Unlike the $N=0.5$ geometry, $\beta^{(1)}$ and $\beta^{(2)}$ show an unbalanced power distribution even at the small value of $\kappa=0.02$. The splitting of these two modes is thresholdless $\kappa_{th}=0$, and for any non-Hermiticity parameter greater than zero the power of these modes will be unequally distributed in the gain and loss sections. Two additional modes ($\beta^{(0)}$ and $\beta^{(3)}$) reach the broken phase, but these modes possess an EP. One can see the power is amplified and attenuated for $\beta^{(1)}$ and $\beta^{(2)}$ in Fig.~\ref{fig6}(c), while all other modes show no variation in magnitude along the propagation direction for $\kappa=0.02$. At $\kappa=0.2$, $\beta^{(1\textendash4)}$ are all in the broken phase and exhibit changes in mode power along the direction of propagation in accordance with the localization of intensity in the gain or loss quadrants [Fig.~\ref{fig6}(d)].

\section{degenerate perturbation theory}

\begin{figure}
\includegraphics[scale=1]{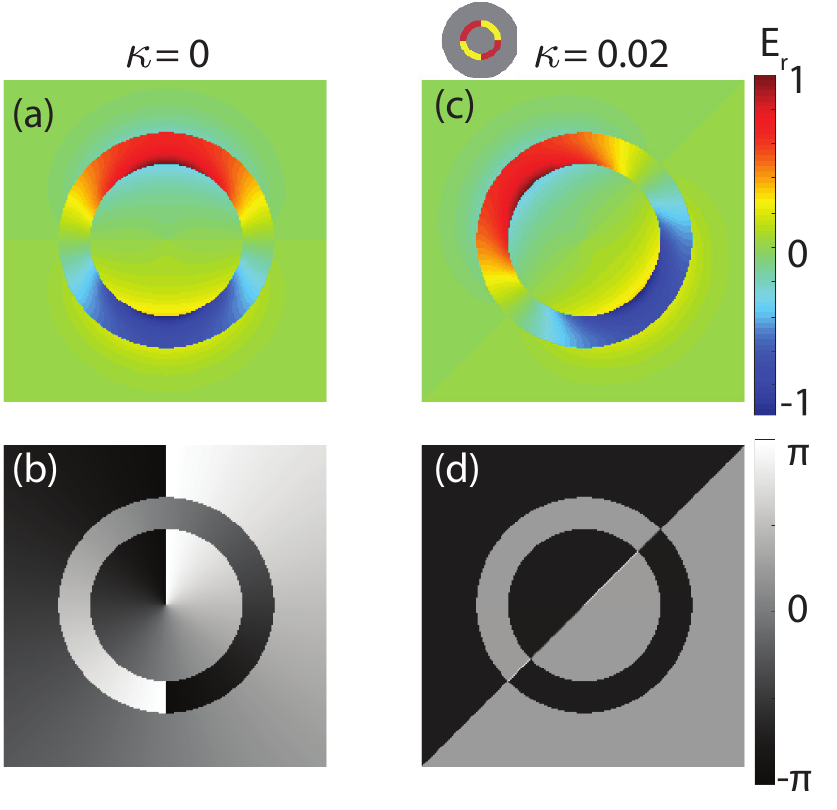}
\caption{\label{fig7} (a) $E_r$ for $m=1$ when $\kappa=0$. (b) Phase angle of $E_r$ for $m=1$ showing OAM. (c) and (d) show $E_r$ and its phase for $\kappa=0.02$ for the $N=1$ waveguide. No OAM is present. }
\end{figure}

As discussed in the previous section, the modes of the coaxial waveguide have 2-fold degeneracy for any $m\ne 0$. In other words, both $\pm m$ modes have the same propagation constants. In this section, we use first-order degenerate perturbation theory to investigate the effect of a $\mathcal{PT}$ symmetric perturbation on the waveguide modes and their propagation constants. This analysis provides a fundamental explanation for the thresholdless behavior of the $N=1$ waveguide and the absence of this behavior in the $N=0.5$ waveguide. More specifically, it provides the sufficient conditions for thresholdless transitions which we extend to larger values of $N$. In the degenerate basis of the $\ket{\pm m}$, the perturbing Hamiltonian $i\hat{H}_I$ takes the following form: 

\begin{equation}
i\begin{bmatrix}
\braket{+m|\hat{H}_I|+m} && \braket{+m|\hat{H}_I|-m}\\
\braket{-m|\hat{H}_I|+m} && \braket{-m|\hat{H}_I|-m}
\end{bmatrix}.
\end{equation}
Due to symmetry, $\braket{+m|\hat{H}_I|+m} = \braket{-m|\hat{H}_I|-m}$ and $\braket{-m|\hat{H}_I|+m} = \braket{+m|\hat{H}_I|-m}^*$. However, from Eq.~\ref{HI-matrix2}, we have:

\begin{equation}
H_{I_{m,m}}=H_{I_{-m,-m}}=0
\end{equation}
and
\begin{equation}
H_{I_{m,-m}}=H_{I_{-m,m}}^*=\alpha_m = 2\pi\omega\epsilon_0C_{2m}\int\limits_{R_{in}}^{R_{out}}drr|R_m(r)|^2 .
\end{equation}
Accordingly, the perturbation matrix simplifies to:

\begin{equation}
\begin{bmatrix}
0 && i\alpha_m\\
i\alpha_m^* && 0
\end{bmatrix},
\end{equation}
and the new eigenvalues (i.e., propagation constants) of the perturbed $m^{th}$ states are:
\begin{equation}\label{G/L modes}
\beta_{L,G}=\beta_m\pm i|\alpha_m|.
\end{equation}
Note that $\alpha_m$ is linearly proportional to $C_{2m}$ and thus $\kappa$. Therefore if $C_{2m}\ne 0$, the degeneracy of $\ket{\pm m}$ is lifted and the eigenvalues move into the complex plane where exclusive gain and loss modes are obtained ($\kappa_{th}=0$). Using the properties of the Fourier coefficients, the following relations for the eigenstates of the loss and gain modes are derived:

\begin{equation}\label{eigenstateloss}
\text{loss eigenstate} \rightarrow\frac{1}{\sqrt{2}}(1,-i)
\end{equation}

\begin{equation}\label{eigenstategain}
\text{gain eigenstate} \rightarrow\frac{1}{\sqrt{2}}(1,+i).
\end{equation}
These coefficients are the weights of the new modes in the basis of the CCW ($\ket{+m}$) and CW ($\ket{-m}$) modes. The sign difference in the imaginary component reveals the modes are $\pi/2$ out of phase with one another. This combination is unique compared to the $\kappa=0$ modes, despite the similarity in the field profiles at first glance. The $m=1$ electric field for the $\kappa=0$ waveguide is plotted in Fig.~\ref{fig7}(a), while Fig.~\ref{fig7}(b) shows the phase. The sweep from $-\pi \rightarrow \pi$ shows the OAM of this mode is $m=1$. The electric field of one of the split degenerate modes for $N=1$ is included as Fig.~\ref{fig7}(c) for comparison. Although the mode appears to be the product of a simple rotation, the phase angle in Fig.~\ref{fig7}(d) confirms the new mode is no longer a pure OAM mode and does not rotate as it propagates. 

As seen in Fig.~\ref{Fig1}, $C_{2m}$ always vanishes for half integer values of $N$. Therefore, to break the degeneracy of the CW and CCW modes and enter the broken phase with $\kappa_{th}=0$, $N$ must be an integer. This conclusion is consistent with Ref.~\citenum{Feng:2014gg}, where the degeneracy between the azimuthal modes was broken for a properly chosen number of gain/loss sections. Based on Eq.~\ref{G/L modes}, the new eigenvalues have the same real part (propagation constant), but one of the modes is amplified as it propagates while the other is attenuated. The broken phase is entered without an EP, and the imaginary component of the propagation constant is linear with $\kappa$.

The predictions of degenerate perturbation theory are all confirmed with our numerical analysis from the previous sections. In Fig.~\ref{Fig1}(b), the $N=1$ structure has $C_2\ne 0$, and the degeneracy of the $\pm1$-modes ($\beta^{(1)}$ and $\beta^{(2)}$) is lifted for $\kappa\ne0$. The linear dependence of the imaginary component of the propagation constant with $\kappa$, plotted in Fig.~\ref{fig5}(b), is numerically found to be 0.026 nm$^{-1}$, which is in very good agreement with the slope predicted by degenerate perturbation theory analysis, 0.027 nm$^{-1}$. 

The localization of the field intensity, in addition to an amplification or absorption of a given orientation, suggests a segment of coaxial waveguide with a $N=1$ configuration of gain and loss would be able to convert an input mode with one unit of angular momentum, e.g., $+1$ or $-1$, into an azimuthally localized mode. Two different $N=1$ coaxial segments with different orientations could multiplex two orthogonal gain modes into a single passive coaxial waveguide. The same $N=1$ coaxial waveguide design could then demultiplex the signals by absorbing one orientation while allowing the other to pass though. 

\begin{figure}
\includegraphics[scale=1]{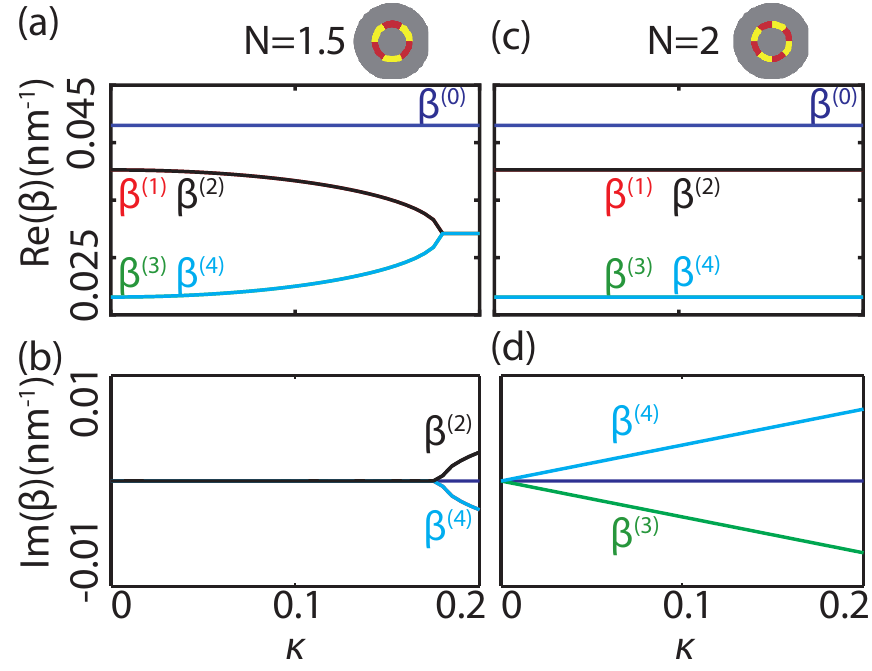}
\caption{\label{fig8} Real (a) and imaginary (b) parts of the wavevectors of the five lowest order modes of a $\mathcal{PT}$ symmetric waveguide for $N=1.5$. Real (c) and imaginary (d) parts of the wavevectors of the five lowest order modes of a $\mathcal{PT}$ symmetric waveguide for $N=2$.}
\end{figure}

Figure~\ref{fig8} shows the modal properties of two other cases of $\mathcal{PT}$ symmetric waveguides, where $N=1.5$ (a,b) and $N=2$ (c,d). While $N=1.5$ has the signature of typical phase breaking (i.e., non-zero threshold), the $N=2$ waveguide possesses thresholdless symmetry breaking. As seen in Fig.~\ref{Fig1}(b), $C_4$ is the first non-zero Fourier coefficient of $N=2$. Hence, $N=2$ can remove the degeneracy of the $\pm 2$-modes ($\beta^{(3)}$ and $\beta^{(4)}$), but not the $\pm 1$-modes. Although the real parts of the propagation constants for $N=2$ show no variation with $\kappa$ in Fig.~\ref{fig8}(c), the imaginary parts of the $\beta$s in Fig.~\ref{fig8}(d) do change. While $\beta^{(0)},\beta^{(1)}$, and $\beta^{(2)}$ remain constant and lossless, $\beta^{(3)}$ and $\beta^{(4)}$ avoid crossing in the imaginary plane with a linear dependency. Through a linear fitting, the slope of this line is determined to be 0.035 nm$^{-1}$, which is in good agreement with 0.034 nm$^{-1}$ predicted by perturbation analysis ($H_{I_{2,-2}}$ perturbing matrix element). This linear dependency is indicative of the splitting described by Eq.~\ref{G/L modes}.

\section{conclusion}

In conclusion, we have proposed and studied a $\mathcal{PT}$ symmetric coaxial waveguide with varying numbers of paired gain and loss sections arranged azimuthally within the dielectric channel. If the gain and loss do not possess inversion symmetry, the eigenvalues of the waveguide possess EPs. Conversely, distributions of gain and loss with inversion symmetry are shown to exhibit thresholdless $\mathcal{PT}$ symmetry breaking for degenerate modes that match the periodicity of the gain and loss, e.g., $N=1$ drives the $m=\pm1$ modes to become amplifying and attenuating complex conjugates. Degenerate perturbation theory confirms the distribution of gain and loss, as defined through Fourier coefficients, determines if thresholdless behavior is achievable for a given structure. These structures transform a degenerate mode pair into complex conjugates according to a relation with the Fourier coefficients:  if $C_{2m}\ne0$, then $\ket{\pm m}$ degeneracy is lifted. In an unperturbed coaxial waveguide, these degenerate modes represent different OAM states, CW and CCW. However, when the degeneracy is lifted, the newly formed modes are equal combinations of phase offset CW and CCW, and therefore have no OAM. 

Given that these degenerate and complex conjugate broken-symmetry modes all possess the same propagation constant, small cross-section plasmonic coaxial waveguides could be designed with MDM that use both modes to double data throughput. An input OAM input would be converted in a $N=1,$ $\mathcal{PT}$ symmetric, filtering waveguide to a new hybrid mode with a specific spatial distribution of field intensity. This new hybrid mode could be transferred to a $\kappa=0$ plasmonic channel and multiplexed with the other spatial distribution. These multiplexed signals could then be demultiplexed by splitting the coaxial waveguide into two new $N=1$, $\mathcal{PT}$ symmetric, filtering waveguides that would absorb all of the misaligned spatially localized information while amplifying the aligned signal. The metallic cladding present in the coaxial design also allows for dense packing of the waveguides with minimal cross talk and interference, allowing these different multiplexing and demultiplexing waveguides to be closely spaced. If a design can support multiple wavevectors, additional degenerate pairs (e.g., $\ket{\pm 2}$) can be used to further multiply throughput.

\section*{Acknowledgements}
The authors appreciate all Dionne group members for their insightful feedback on the work. Funding from a Presidential Early Career Award administered through the Air Force Office of Scientific Research (grant no. FA9550-15-1-0006) is gratefully acknowledged.
\appendix

\section{IMPERFECTION EFFECTS FROM METALLIC LOSS}

\begin{figure}
\includegraphics[scale=1]{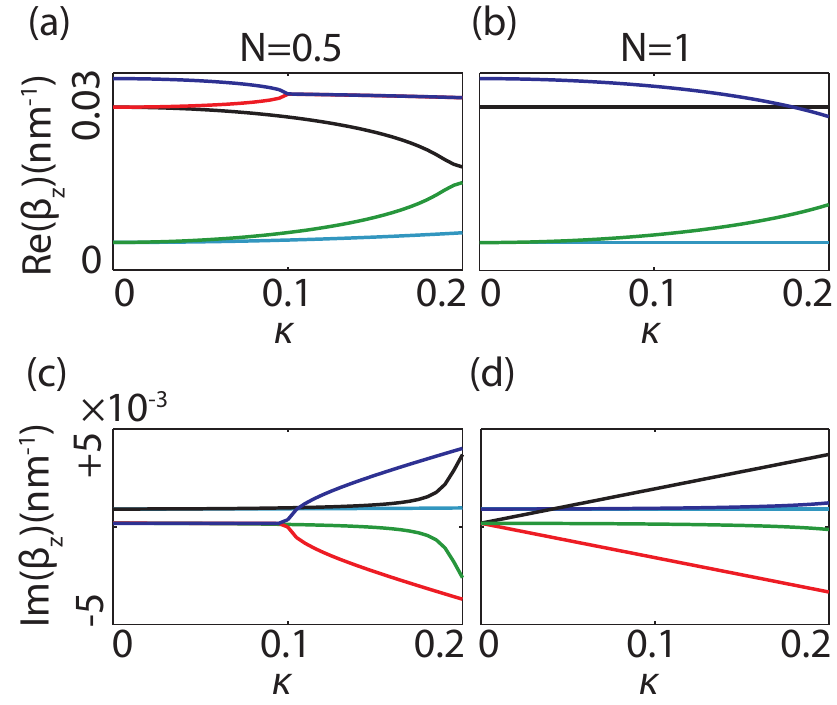}
\caption{\label{fig9} Modes of the MIM coaxial waveguide as a function of $\kappa$ (real,imaginary) for $N=0.5$ (a,c) and $N=1$ (b,d) at $E=2$\,eV. The empirical JC dataset has been used for this set of calculations.}
\end{figure}

In all previous sections, the calculations have been performed assuming a lossless metal. \footnote{In the presence of the metal loss the modes of the uniform waveguide do not make a complete basis any more. However the power orthogonality condition is still held.} This assumption is hard to achieve for plasmonic materials at optical wavelengths; therefore it is necessary to investigate how the predicted EPs and trends change as the metal losses are included. 

In this appendix, we present the results of the 3-layer coaxial waveguide with a $\mathcal{PT}$ symmetrically modulated channel when the permittivity of Ag is given by Johnson-Christy empirical data.\cite{PhysRevB.6.4370} The results presented in Fig.~\ref{fig9} include this empirical data and can be compared to Fig.~\ref{Fig3} and Fig.~\ref{fig5} from the main text.

Due to the difference of the bulk plasma frequency of the empirical data and the Drude model, the range of the values are slightly different. The main difference is the shift of the EP to larger values of $\kappa$ for both cases of $N=0.5$ and $N=1$. However, aside from the difference in the range, the trends are similar. As seen in Fig.~\ref{fig9}(a), for $N=0.5$ there is a clear EP for ($\beta^{(0)}$ and $\beta^{(1)}$) as it was for the lossless case. Since the modes are lossy at $\kappa=0$, the imaginary part of the propagation constants are non-zero even before the EP [Fig.~\ref{fig9}(c)]. However, when $\kappa$ reaches the threshold value, the imaginary parts avoid crossing and diverge symmetrically. 

The real and imaginary parts of the propagation constants for $N=1$ are shown in Fig.~\ref{fig9}(b) and(d), respectively. Similar to the lossless case, the real parts of $m=\pm 1$ ($\beta^{(1)}$ and $\beta^{(2)}$) remain the same as $\kappa$ varies. However, their imaginary parts separate from each other right after $\kappa=0$ and vary linearly as a function of the non-Hermiticity parameter [Fig.~\ref{fig9}(d)]. In contrast to the lossless case, these modes are not purely gain or loss modes, and there is an offset due to the loss of the uniform coaxial modes. Nevertheless, this offset only introduces a shift in the values of the imaginary part and it does not change the linear trends of the branches. In our specific geometry, the pair of these modes reach Im($\beta$)$=0$ at a very low values of $\kappa=0.01$.

\section{EIGENVALUE PROPERTIES}
When expanded in the basis of the non-perturbed uniform coaxial waveguide modes, the problem of modal distributions and dispersions is simplified to the following eigen-equation for vector of expansion coefficients $V$ and propagation constants of the hybrid modes $\beta$:
\begin{equation}\label{eigenvalue}
(H_R+iH_I)\ket{V}=\beta \ket{V}
\end{equation}

Note that $\hat{H}_R$ and $\hat{H}_I$ are both Hermitian operators. In this Appendix, we find the general properties of the eigenvalues of Eq.~\ref{eigenvalue} in the complex plane. Specifically, we will show that $\beta$ appear either as real values or complex conjugate pairs.

Modal propagation constants are the solutions of the following secular equation:
\begin{equation}
det(H_R+iH_I-\beta I)=0
\end{equation}
where $H_R$ is a diagonal matrix with real entries, and $H_I$ is a Hermitian matrix with zero diagonal elements. Taking the complex conjugate of the above equation gives:
\begin{equation}
det(H_R^*-iH_I^*-\beta^* I)=0.
\end{equation}
However, $H_R^*=H_R=H_R^T$ and $H_I^*=H_I^T$; hence, the above equation can be rewritten as:
\begin{equation}
det((H_R-iH_I)^T-\beta^* I)=0.
\end{equation}
Therefore, $\beta$ and $\beta^*$ are the eigenvalues of $H_R+iH_I$ and $(H_R-iH_I)^T$, respectively. The latter implies that $\beta^*$ is an eigenvalue of $(H_R-iH_I)$ as well. $(H_R-iH_I)$ is the effective matrix of a $\mathcal{PT}$ symmetric waveguide when the gain and loss sectors are swapped. While this rotation changes the eigenvectors of the matrix in Eq.~\ref{eigenvalue}, it does not change the eigenvalues, which means $\beta^*$ is also an eigenvalue of Eq.~\ref{eigenvalue}. Accordingly, the propagation constants of a $\mathcal{PT}$ symmetric waveguide are either real values or complex conjugate pairs. 

\section{EXPANSION COEFFICIENTS OF THE $\mathcal{PT}$ SYMMETRIC MODES IN THE BASIS OF THE UNIFORM COAXIAL WAVEGUIDE}

\begin{figure}
\includegraphics[scale=1]{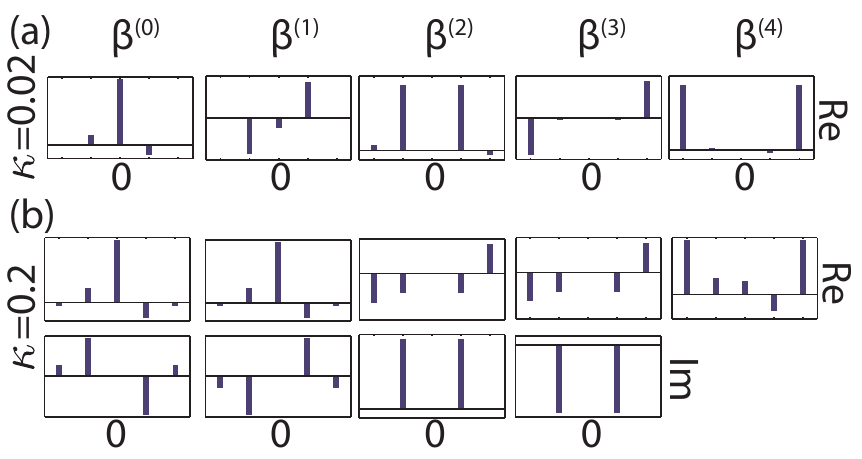}
\caption{\label{fig10} Expansion coefficients of each mode for $N=0.5$ at $E=2$\,eV. Coefficients are plotted for the same five modes shown in Fig.~\ref{Fig3} at two values of the non-Hermiticity parameter. Since the coefficients are complex in general, we have plotted the real and imaginary parts separately. In each panel, the continuous horizontal line is 0.}
\end{figure}

\begin{figure}
\includegraphics[scale=1]{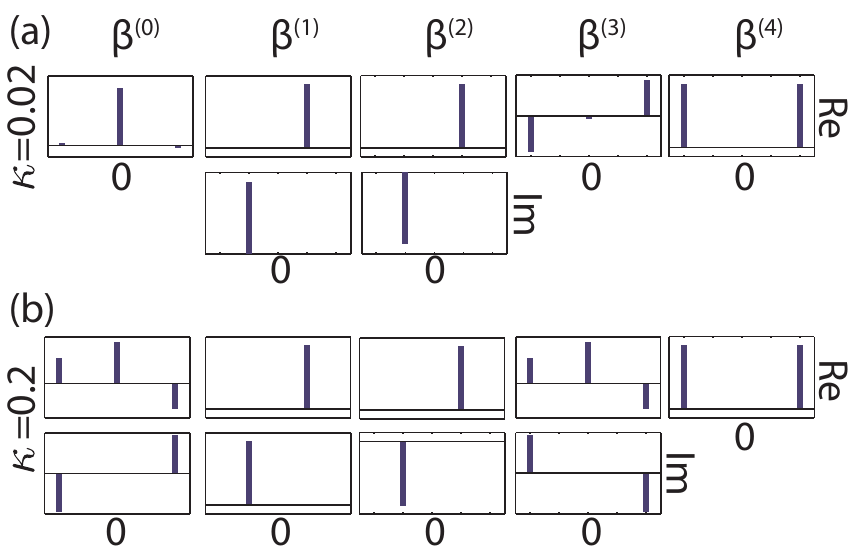}
\caption{\label{fig11} Expansion coefficients of each mode for $N=1$ at $E=2$\,eV. Coefficients are plotted for the same five modes shown in Fig.~\ref{fig5} at two values of the non-Hermiticity parameter. Since the coefficients are complex in general, we have plotted the real and imaginary parts separately. In each panel, the continuous horizontal line is 0.}
\end{figure}

In the main text, we discussed the properties of the eigenvalues of the $\mathcal{PT}$ symmetric waveguides, i.e., propagation constant of the modes, and showed how the morphology of the modes can be controlled via the number of gain/loss sections. This appendix extends the existing argument to the properties of the eigenmodes. It discusses how the modes of the new system are related to the modes of the uniform waveguide. Figure~\ref{fig10} shows the expansion coefficients of the five lowest order modes for $N=0.5$. Figure~\ref{fig11} shows the expansion coefficients of the five lowest order modes for $N=1$. The weightings of the coefficients for $\beta^{(1)}$ and $\beta^{(2)}$ confirm the description of the eigenstates in Eq.~\ref{eigenstateloss} and Eq.~\ref{eigenstategain}. Note that due to the mixing of the modes with different angular momenta, the new modes have a different beating pattern around the ring as shown in Fig.~\ref{Fig3}(c,d) and Fig.~\ref{fig5}(c,d). This beating pattern is due to the interference of the CW and CCW modes within the ring. 

The general behavior of the modes in this case can be directly compared to azimuthal photonic crystals, where the real part of the refractive index is modulated periodically. At the band edge of these photonic crystals, the two modes that exist will be preferentially confined in one type of material. Since the variation is in the real part of the refractive index, these spatial distributions would lead to a change in the energy of the modes (i.e., real part of the propagation constant). Unlike the photonic crystal case, here the variation happens for the imaginary part of the refractive index. The $\mathcal{PT}$ symmetry breaking occurs at a point where the modes are dominantly in the loss or gain sections. However, this spatial distribution does not change the phase velocity and only changes the propagation length of the modes.

\bibliography{PT-MDM-final}
\end{document}